\documentclass[sigconf,screen,authorversion,nonacm]{acmart}

\AtBeginDocument{%
  }

\setcopyright{acmcopyright}
\copyrightyear{2018}
\acmYear{2018}
\acmDOI{XXXXXXX.XXXXXXX}

\acmConference[ICSE]{}{April 14-20, 2024}{Lisbon, Portugal}
\acmPrice{}
\acmISBN{}

\usepackage{booktabs} 
\usepackage{siunitx} 
\usepackage{tabularx} 
\usepackage{xspace}
\usepackage{multirow}
\usepackage{listings}
\usepackage{enumitem}
\usepackage{tikz}
\usepackage{colortbl}

\definecolor{codegreen}{rgb}{0,0.6,0}
\definecolor{codegray}{rgb}{0.5,0.5,0.5}
\definecolor{codepurple}{rgb}{0.58,0,0.82}
\definecolor{backcolour}{rgb}{0.95,0.95,0.92}

\lstdefinestyle{pstyle}{
	commentstyle=\color{codegreen},
	keywordstyle=\color{codegreen},
	numberstyle=\tiny\color{codegray},
	stringstyle=\color{red},
	basicstyle=\ttfamily\fontsize{8}{8}\selectfont,
	breakatwhitespace=false,         
	breaklines=true,                 
	captionpos=b,                    
	keepspaces=true,                 
	numbers=none,                    
	numbersep=5pt,                  
	showspaces=false,                
	showstringspaces=false,
	showtabs=false,                  
	tabsize=4
}

\newcommand*{\priority}[1]{%
	\pgfmathsetmacro\percentage{#1} 
	\begin{tikzpicture}[scale=0.125]%
		\draw (0,0) circle (1);
		\pgfmathparse{\percentage > 50 ? 1 : 0}
		\ifnum\pgfmathresult=1
		\fill[fill opacity=0.5,fill=green!60!black] (0,0) -- (90:1) arc (90:90-\percentage*3.6:1) -- cycle; 
		\else
		\pgfmathparse{\percentage > 25 ? 1 : 0}
		\ifnum\pgfmathresult=1
		\fill[fill opacity=0.5,fill=blue!60!black] (0,0) -- (90:1) arc (90:90-\percentage*3.6:1) -- cycle; 
		\else
		\fill[fill opacity=0.5,fill=red!80!black] (0,0) -- (90:1) arc (90:90-\percentage*3.6:1) -- cycle; 
		\fi
		\fi
	\end{tikzpicture}%
}

\newcommand*{\circlecolor}[1]{%
	\begin{tikzpicture}[scale=0.125]%
		\draw (0,0) circle (1);
		\ifnum#1=1
		\fill[fill opacity=0.7,fill=blue!70!black] (0,0) circle (1);
		\fi
	\end{tikzpicture}%
}

\usepackage[colorinlistoftodos,prependcaption,textsize=tiny]{todonotes}

\newcommand{\toolname}{HeaderGen\xspace}

\newcommand{\typeevalpy}{\textsc{TypeEvalPy}\xspace}

\title[TypeEvalPy: A Micro-benchmarking Framework for Python Type Inference Tools]{TypeEvalPy: A Micro-benchmarking Framework for\\ Python Type Inference Tools}

\author{Ashwin Prasad S. Venkatesh\textsuperscript{§}, 
	Samkutty Sabu\textsuperscript{¶},
	Jiawei Wang\textsuperscript{†}, 
	Amir M. Mir\textsuperscript{‡},
	Li Li\textsuperscript{*},
	Eric Bodden\textsuperscript{**}}
\affiliation{%
	\textsuperscript{§}\textit{ashwin.prasad@upb.de}, Heinz Nixdorf Institut, Paderborn University, Paderborn, Germany\\
	\textsuperscript{¶}\textit{samkutty@mail.uni-paderborn.de}, Paderborn University, Paderborn, Germany\\
	\textsuperscript{†}\textit{jiawei.wang1@monash.edu}, Faculty of Information Technology, Monash University, Melbourne, Australia\\
	\textsuperscript{‡}\textit{s.a.m.mir@tudelft.nl}, Delft University of Technology, Delft, The Netherlands\\
	\textsuperscript{*}\textit{lilicoding@ieee.org}, School of Software, Beihang University, Beijing, China\\
	\textsuperscript{**}\textit{eric.bodden@upb.de}, Heinz Nixdorf Institut \& Fraunhofer IEM, Paderborn University, Paderborn, Germany\\
	\country{}
}

\renewcommand{\shortauthors}{Venkatesh et al.}

\begin{document}

%
%
%
%
%
%

\begin{abstract}
In light of the growing interest in type inference research for Python, both researchers and practitioners require a standardized process to assess the performance of various type inference techniques. This paper introduces \typeevalpy, a comprehensive micro benchmarking framework for evaluating type inference tools. \typeevalpy contains 154 code snippets with 845 type annotations across 18 categories that target various Python features.
The framework manages the execution of containerized tools, transforms inferred types into a standardized format, and produces meaningful metrics for assessment.
Through our analysis, we compare the performance of six type inference tools, highlighting their strengths and limitations.
Our findings provide a foundation for further research and optimization in the domain of Python type inference.
\end{abstract}

%



\maketitle

\section{Introduction}
Type inference refers to the process of automatically determining the data type of an expression within a programming language.
In Python, which is dynamically typed, this determination takes place at runtime.
To address potential ambiguities, developers can utilize type annotations, which explicitly specifies the expected data types of variables or function returns.
As the complexity of software projects increases, programmers find it increasingly challenging to maintain consistent data types.
In response to this challenge, both industry and academia have developed type inference tools and static type checkers.
Examples from industry include \emph{Pyright}~\cite{pyright} and \emph{Pytype}~\cite{Pytype2023a}, while academic contributions feature \emph{Type4Py}~\cite{type4py} and \emph{HiTyper}~\cite{HiTyper}.
The topic of type inference in Python is a growing area of research and tool development within the software engineering community. 
Current efforts focus on understanding the advantages of enforcing type annotations, and on finding ways to infer types in Python code that lacks type annotations.

In recent years, many solutions for type inference have been proposed. 
However, a unified and comprehensive evaluation framework for these tools is still lacking.
Current literature primarily assesses the performance of such type inference tools based on large-scale real-world benchmark datasets, notably \emph{Type4Py}, \emph{HiTyper}, and \emph{Typilus} \cite{Typilus}.
On the contrary, open-sourced solutions only rely on specifically-designed test cases.
This evaluation approach, nonetheless, presents several limitations:
(1) Different studies might report findings based on different datasets, complicating a direct comparison and understanding of the relative merits and drawbacks of each tool.
(2) Type annotations in real-world datasets are sometimes erroneous. 
(3) Evaluations often provide a broad-brush score, overlooking nuanced insights into specific technical challenges, including the treatment of diverse language constructs.

In this paper, we introduce \typeevalpy, a type inference evaluation framework for Python bundled with a micro-benchmark that covers all the Python language constructs of Python 3.10.

Our primary objective with \typeevalpy is to provide insights into the recent advances in type inference tooling for Python programs.
When presented with an executable type inference tool, \typeevalpy processes the tool with input from the built-in micro-benchmark and outputs the inferred type information in a standard format for further analysis.
Then, \typeevalpy analyzes the output of each tool and reports the comparative analysis using a set of metrics such as exact match rate, precision, etc (c.f. section~\ref{sec:analyzer}). 

We demonstrate \typeevalpy's utility by evaluating six state-of-the-art type inference tools, including two ML-based and four static analysis-based approaches.
For ML-based approaches, we extend our analysis to incorporate top-$n$ predictions.
Our empirical findings reveal that the performance of type inference can be significantly enhanced by integrating external user-annotated type stubs and combining static analysis with ML techniques.
The state-of-the-art hybrid strategy in \emph{HiTyper}, outperforms its purely static analysis based counterpart.
However, the underlying static analysis technique that \emph{HiTyper} uses, performs poorly compared to the other pure static analysis-based alternatives, indicating that the performance of \emph{HiTyper} can be improved.
Moreover, we posit that researchers should place greater emphasis on function parameter annotations, particularly since the majority of tools generate only a limited number of these annotations.
Nevertheless, achieving soundness in type inference remains a challenge, with even the top-performing tools achieving a soundness rate of only 44\%.

The paper is organized as follows: The technical design is detailed in section~\ref{sec:typeevalpy} followed by experimental results being reported in section~\ref{sec:experiments}. 
After a brief discussion about the results in section~\ref{sec:discussion}, the paper is summarized in section~\ref{sec:conclusion}. 

\textbf{Availability.}
\typeevalpy is published on GitHub as open-source software: \url{https://github.com/secure-software-engineering/TypeEvalPy}

\section{TypeEvalPy Framework}
\label{sec:typeevalpy}

The primary goal of \typeevalpy is to offer a comprehensive, standardized, and reproducible benchmarking system for evaluating type inference tools in Python.
To this end, \typeevalpy contains a diverse set of 154 code snippets with 845 type annotations across 18 categories that capture the nuances of various Python features.
For a tool to be benchmarked using \typeevalpy, it must be adapted into a containerized format that aligns with \typeevalpy's specifications. 
To simplify this adaptation process, we are sharing a template. 
For reference, \typeevalpy already includes containerized versions of six type inference tools: \emph{HeaderGen}~\cite{Headergen}, \emph{Jedi}~\cite{jedi}, \emph{pyright}~\cite{pyright}, \emph{Scalpel}~\cite{li2022scalpel}, \emph{HiTyper}~\cite{HiTyper}, and \emph{Type4Py}~\cite{type4py}.

The \typeevalpy framework is organized into three main modules: \texttt{Runner}, \texttt{Translator}, and \texttt{Result Analyzer}.
First, the \texttt{runner} module manages the initiation and execution of containerized tools, specifically running type inference on the micro-benchmark.
Following this, the \texttt{translator} module takes on the role of transforming the inferred types into a standardized format, making them suitable for comparison.
Lastly, the \texttt{analyzer} compiles the results from all the tools and provides meaningful metrics for assessment.
Importantly, the modules realize a high degree of automation, thereby ensuring the reproducibility of results in the academic context.

\subsection{Micro-benchmark}


The micro-benchmark of \typeevalpy comprises of 154 program snippets containing 845 type annotations.
To ensure a comprehensive coverage of Python's language features, snippets are subdivided into 18 categories, each representing a specific feature.
This categorization is based on PyCG's \cite{PyCG} call graph benchmark and we extend it to address gaps in the coverage of language features.
Our strategy for enhancing the benchmark involved inspecting the Python manual to ensure all language features are adequately represented.
Furthermore, we categorized test cases from existing inference tools, ensuring the comprehensiveness of our benchmark. Additionally, the micro-benchmark's modular design allows for easy extensions.
The micro-benchmark consists of the following categories:
\emph{args (8)}, \emph{assignments (8)}, \emph{builtins (7)}, \emph{classes (26)}, \emph{decorators (8)}, \emph{dicts (15)}, \emph{direct\_calls (6)}, \emph{dynamic (3)}, \emph{exceptions (2)}, \emph{external (7)}, \emph{functions (9)}, \emph{generators (6)}, \emph{imports (14)}, \emph{kwargs (4)}, \emph{lambdas (6)}, \emph{lists (10)}, \emph{mro (7)}, \emph{returns (8)}.

Type annotation format is based on the Scalpel framework~\cite{li2022scalpel} and are stored as JSON files with the code snippets.
They contain the following:
(1) \textit{file}: denotes the filename,
(2) \textit{line\_number:} specific line in the file.
(3) \textit{col\_offset:} represents the indentation.
(4) \textit{type:} list of types.
(5) \textit{function:} function name, if the annotation is within one.
(6) \textit{variable:} specifies the variable's name being annotated.
(7)~\textit{parameter:} captures the name of a function argument, if relevant. 

Type annotations are categorized into three categories: (1) Function return (FR) type, (2) Function parameter (FP) type, and (3) Local variable (LV) type.
In total, the micro-benchmark consists of 239 FR, 88 FP, and 518 LV type annotations.
To construct the ground truth, the first two authors manually inspected each code snippet and, where required, used a debugger to verify the run-time type of each Python element.
To further mitigate potential errors, each file was reviewed consecutively by both authors.

During the development of the ground truth, we made several design decisions:
(1) Type annotations for generics are not concretized, for instance, a list of integers is annotated as \texttt{List} instead of \texttt{List[Int]}.
(2) FPs are annotated based on their usage. In cases where a function can return multiple types, special care was taken to ensure that the function is called with parameters of all types the function supports.
(3) FR types and local variables defined inside a function are context-insensitive, i.e., if a variable defined inside a function can take on multiple types based on different calling contexts, each variable is annotated with all possible types it can have during runtime.
(4) More generally, types assigned to entities were chosen to reflect all the possible runtime types \emph{in the given program}, and we chose the most specific type possible. Therefore, none of the type annotations are marked as ``\texttt{Any}''.

\subsection{Runner and Translator}

The primary responsibility of the \texttt{runner} module is to orchestrate the execution of containerized type inference tools on the micro-benchmark. 
For each type inference tool, the \texttt{runner} module creates an instance of the Docker container, the micro-benchmark is copied into the running instance and runs the type inference inside it.
Then, the \texttt{runner} module uses the \texttt{translator} to convert results into the \typeevalpy format.
Once each tool finishes running, the \texttt{runner} module calls the \texttt{result analyser} module.


\subsection{Result Analyzer}
\label{sec:analyzer}
The \texttt{analyzer} module produces detailed statistics for comparing the effectiveness of different tools as listed below:

\begin{itemize}[leftmargin=.3cm]
    \item \textbf{Exact matches:} The number of inferred types that exactly match the ground truth. This metric is used widely used in the literature to evaluate type inference tools  \cite{Typilus,HiTyper,type4py}.
	\item \textbf{Precision:} The fraction of reported types that are exactly inferred according to the ground truth.
	\item \textbf{Recall:} The number of actual types that are exactly reported by the type inference tool.
	\item \textbf{Soundness:} Whether the type inference tool identifies all possible types specified in the Python code to ensure none are omitted. Reported as a boolean per code snippet in the micro-benchmark.
	\item \textbf{Completeness:} Whether the tool accurately reports only the types that are present, avoiding any incorrect or extraneous types. Reported as a boolean per code snippet in the micro-benchmark.
	\item \textbf{Top-$n$ prediction comparison:} The accuracy comparison of probabilistic tools when considering their top-$n$ inferred types. This metric is widely used to evaluate ML predictors \cite{Pyinfer,HiTyper,type4py}.
	\item \textbf{Report of missing types:} List of types that are present in the ground truth but are unreported by the tools.
	\item \textbf{Report of mismatched types:} List of types reported by the tools that do not align exactly with the ground truth. 
\end{itemize}

\section{Experiments}
\label{sec:experiments}

\begin{table*}[ht]
	\caption{Comparison of exact matches, sound, and complete values of type inference tools for micro-benchmark categories}
	\label{tab:typeevalpy-table-cats}
	\setlength{\tabcolsep}{4.8pt} 
	\renewcommand{\arraystretch}{.9}
	
	\begin{tabular}{l*{7}{|rrr}}
		\multicolumn{22}{r}{\small  \textbf{FR:} Function return type, \textbf{FP:} Function parameter type, \textbf{LV:} Local variable type, \textbf{845:} Total type annotations, \textbf{154:} Total test cases}   \\
		\toprule
		\multicolumn{1}{c|}{\multirow{2}{*}{\textbf{Category}}} & \multicolumn{3}{c}{\textbf{HeaderGen}} & \multicolumn{3}{c}{\textbf{Jedi}} & \multicolumn{3}{c}{\textbf{Pyright}} & \multicolumn{3}{c}{\textbf{HiTyper-DL}} & \multicolumn{3}{c}{\textbf{HiTyper}} & \multicolumn{3}{c}{\textbf{Scalpel}} & \multicolumn{3}{c}{\textbf{Type4Py}} \\
		\cmidrule(lr){2-4} \cmidrule(lr){5-7} \cmidrule(lr){8-10} \cmidrule(lr){11-13} \cmidrule(lr){14-16} \cmidrule(lr){17-19} \cmidrule(lr){20-22}
		& FR & FP & LV & FR & FP & LV & FR & FP & LV & FR & FP & LV & FR & FP & LV & FR & FP & LV & FR & FP & LV \\
		\midrule
		\rowcolor{black!5}args                                                   & 17                     & 9                      & 12                     & 12                     & 0                      & 9                      & 8                      & 1                      & 8                      & 12                     & 0                      & 6                      & 8                      & 0                      & 0                      & 8                      & 7                      & 0                      & 11                     & 2                      & 6                      \\
		assignments                                            & 15                     & 1                      & 33                     & 20                     & 0                      & 21                     & 20                     & 0                      & 25                     & 21                     & 4                      & 9                      & 20                     & 0                      & 5                      & 20                     & 2                      & 1                      & 0                      & 2                      & 5                      \\
		\rowcolor{black!5} \textbf{builtins}                                               & 0                      & 0                      & 26                     & 0                      & 0                      & 21                     & 1                      & 0                      & 45                     & 1                      & 2                      & 18                     & 1                      & 0                      & 17                     & 0                      & 0                      & 0                      & 1                      & 2                      & 8                      \\
		classes                                                & 39                     & 7                      & 67                     & 0                      & 0                      & 57                     & 1                      & 0                      & 46                     & 27                     & 2                      & 41                     & 24                     & 0                      & 23                     & 25                     & 0                      & 0                      & 0                      & 0                      & 17                     \\
		\rowcolor{black!5}decorators                                             & 11                     & 6                      & 2                      & 10                     & 0                      & 8                      & 7                      & 0                      & 3                      & 8                      & 0                      & 3                      & 7                      & 0                      & 0                      & 16                     & 3                      & 0                      & 7                      & 0                      & 3                      \\
		dicts                                                  & 23                     & 3                      & 60                     & 21                     & 0                      & 34                     & 19                     & 2                      & 50                     & 20                     & 3                      & 22                     & 20                     & 2                      & 16                     & 19                     & 2                      & 1                      & 2                      & 3                      & 18                     \\
		\rowcolor{black!5}direct\_calls                                          & 10                     & 3                      & 8                      & 6                      & 0                      & 7                      & 3                      & 0                      & 6                      & 3                      & 2                      & 4                      & 2                      & 0                      & 0                      & 5                      & 1                      & 0                      & 2                      & 2                      & 4                      \\
		dynamic                                                & 1                      & 0                      & 2                      & 1                      & 0                      & 2                      & 1                      & 0                      & 2                      & 1                      & 0                      & 5                      & 1                      & 0                      & 2                      & 1                      & 0                      & 0                      & 0                      & 0                      & 5                      \\
		\rowcolor{black!5}exceptions                                             & 0                      & 0                      & 2                      & 0                      & 0                      & 1                      & 0                      & 0                      & 1                      & 0                      & 0                      & 1                      & 0                      & 0                      & 1                      & 0                      & 0                      & 0                      & 0                      & 0                      & 0                      \\
		\textbf{external}                                               & 0                      & 0                      & 3                      & 0                      & 0                      & 8                      & 0                      & 0                      & 2                      & 1                      & 0                      & 4                      & 0                      & 0                      & 2                      & 0                      & 1                      & 0                      & 1                      & 0                      & 1                      \\
		\rowcolor{black!5}functions                                              & 8                      & 9                      & 12                     & 5                      & 0                      & 14                     & 5                      & 2                      & 13                     & 5                      & 5                      & 5                      & 3                      & 2                      & 1                      & 6                      & 5                      & 1                      & 1                      & 3                      & 4                      \\
		generators                                             & 9                      & 4                      & 17                     & 5                      & 0                      & 23                     & 4                      & 3                      & 18                     & 10                     & 5                      & 11                     & 10                     & 3                      & 11                     & 6                      & 1                      & 3                      & 1                      & 1                      & 3                      \\
		\rowcolor{black!5}imports                                                & 3                      & 0                      & 11                     & 1                      & 0                      & 16                     & 3                      & 0                      & 20                     & 3                      & 0                      & 11                     & 3                      & 0                      & 0                      & 3                      & 0                      & 0                      & 3                      & 0                      & 10                     \\
		kwargs                                                 & 8                      & 5                      & 5                      & 7                      & 0                      & 4                      & 4                      & 0                      & 5                      & 7                      & 0                      & 0                      & 4                      & 0                      & 0                      & 4                      & 4                      & 0                      & 3                      & 0                      & 0                      \\
		\rowcolor{black!5}lambdas                                                & 3                      & 7                      & 4                      & 6                      & 0                      & 11                     & 2                      & 0                      & 1                      & 3                      & 0                      & 7                      & 3                      & 0                      & 5                      & 2                      & 4                      & 0                      & 0                      & 0                      & 2                      \\
		lists                                                  & 14                     & 1                      & 26                     & 17                     & 0                      & 27                     & 13                     & 0                      & 25                     & 16                     & 3                      & 16                     & 13                     & 0                      & 13                     & 16                     & 1                      & 0                      & 2                      & 3                      & 4                      \\
		\rowcolor{black!5}mro                                                    & 14                     & 0                      & 16                     & 0                      & 0                      & 13                     & 0                      & 0                      & 14                     & 15                     & 0                      & 11                     & 13                     & 0                      & 6                      & 13                     & 0                      & 0                      & 0                      & 0                      & 4                      \\
		returns                                                & 11                     & 1                      & 16                     & 11                     & 0                      & 17                     & 9                      & 0                      & 13                     & 10                     & 1                      & 5                      & 9                      & 0                      & 0                      & 11                     & 1                      & 0                      & 5                      & 1                      & 5                      \\ \midrule
		\multicolumn{1}{c}{\multirow{2}{*}{\textbf{Total}}}   & \textbf{186}           & \textbf{56}            & \textbf{322}           & \textbf{122}           & \textbf{0}             & \textbf{293}           & \textbf{100}           & \textbf{8}             & \textbf{297}           & \textbf{163}           & \textbf{27}            & \textbf{179} & \textbf{141}                    & \textbf{7}                      & \textbf{102}                    & \textbf{155}                    & \textbf{32}                     & \textbf{6}                      & \textbf{39}                     & \textbf{19}                     & \textbf{99}          \\ \cmidrule(lr){2-4} \cmidrule(lr){5-7} \cmidrule(lr){8-10} \cmidrule(lr){11-13} \cmidrule(lr){14-16} \cmidrule(lr){17-19} \cmidrule(lr){20-22}
		\multicolumn{1}{c}{}                                   & \multicolumn{3}{c}{\textbf{564/845} \priority{66} }                                         & \multicolumn{3}{c}{\textbf{415/845} \priority{49}}                                         & \multicolumn{3}{c}{\textbf{405/845} \priority{47}}                                         & \multicolumn{3}{c}{\textbf{369/845} \priority{43}} & \multicolumn{3}{c}{\textbf{250/845} \priority{29}} &\multicolumn{3}{c}{\textbf{193/845} \priority{22}} &\multicolumn{3}{c}{\textbf{157/845} \priority{18}}                                         
		\\ \midrule
		\multicolumn{1}{c}{\textbf{Sound}}                                   & \multicolumn{3}{c}{\textbf{68/154} \priority{44}}                                         & \multicolumn{3}{c}{\textbf{24/154} \priority{15}}                                         & \multicolumn{3}{c}{\textbf{21/154} \priority{13}}                                         & \multicolumn{3}{c}{\textbf{18/154} \priority{11}} & \multicolumn{3}{c}{\textbf{3/154} \priority{0}} &\multicolumn{3}{c}{\textbf{0/154} \priority{0}} &\multicolumn{3}{c}{\textbf{5/154} \priority{0}}                                         
		\\
		\multicolumn{1}{c}{\textbf{Complete}}                                   & \multicolumn{3}{c}{\textbf{55/154} \priority{35}}                                         & \multicolumn{3}{c}{\textbf{30/154} \priority{19}}                                         & \multicolumn{3}{c}{\textbf{91/154} \priority{59}}                                         & \multicolumn{3}{c}{\textbf{32/154} \priority{20}} & \multicolumn{3}{c}{\textbf{135/154} \priority{87}} &\multicolumn{3}{c}{\textbf{81/154} \priority{52}} &\multicolumn{3}{c}{\textbf{11/154} \priority{0}}        \\                                 
		\bottomrule                    	
	\end{tabular}
\end{table*}

We used the \typeevalpy framework to evaluate the most recent versions of the following tools:
HiTyper \cite{HiTyper} and Type4Py \cite{type4py} as examples of machine learning tools,
Jedi \cite{jedi} and Pyright \cite{pyright} were chosen as open-source tools, 
while HeaderGen \cite{Headergen} and Scalpel \cite{li2022scalpel} were picked as representations of academic tools.

HiTyper is a hybrid analysis approach that employs both static analysis and ML to infer types.
It also has an option to only use static analysis for type inference, which is based on PyCG.
We evaluated HiTyper in both modes.
In discussions, the static-analysis method is referred to as ``\textit{HiTyper}," while the hybrid method is referred to as ``\textit{HiTyper-Dl}'', which integrates Type4Py.

\textbf{Results.}
Table~\ref{tab:typeevalpy-table-cats} shows the exact matches of the selected tools for each type categories: 
(1) Function return (FR) type , 
(2) Function parameter (FP) type, and 
(3) Local variables (LV).
The tools are arranged in the table according to their performance from left to right. 
\toolname performed the best with highest overall performance in all categories except in \texttt{builtins} and \texttt{external} categories.
In these specific categories, both Jedi and Pyright performed the best.
This suggests that Jedi and Pyright are more adept at integrating user-specified type hints, commonly referred to as \textit{typestubs}.
On the contrary, Jedi and Pyright fail to infer types of FPs.
Both the tools are designed to infer function parameters as ``\texttt{Any}'' in most cases, except for cases where a function is passed as a parameter.
This strategy, while congruent with Python's duck typing paradigm, limits the applicability of the inferred types in a wider context.  

Among the ML tools, HiTyper-DL outperforms Type4Py with 369 exact matches, while Type4Py had 157 matches.
Additionally, HiTyper-DL shows a notably improved performance compared to HiTyper, which relies solely on static analysis.

\textbf{Soundness and Completeness.} 
Table~\ref{tab:typeevalpy-table-cats} lists the soundness and completeness values at the bottom.
\toolname is sound in 68 of the 154 cases, i.e., it did not miss any types in these cases.
\toolname is complete in 55 of the 154 cases, i.e., it did not falsely identify types in these cases.
\toolname demonstrated the most balanced performance compared to all other tools.
HiTyper has a low score of 3 in soundness, HiTyper-DL improves this score to 18.
While this score is a modest increase, it is notably better than HiTyper.
In the assessment of HiTyper, while its completeness score appears promising, it is noteworthy that the tool did not produce predictions for 34 out of the 154 cases. Furthermore, HiTyper often failed to infer types, as indicated by its soundness and exact matches.

\textbf{top-$n$ Matches.}
Table~\ref{tab:typeevalpy-table-topn_analysis} shows top-$n$ results of ML tools and its comparison with \toolname.
HiTyper-DL shows significant improvement considering top-5 predictions with 441 exact matches which is 78.2\% of \toolname's score.
However, the difference between top-$n$ values of 3 and 5 are small, indicating that the majority of correct predictions by HiTyper-DL fall within the top-3 types.
Type4Py benefits immensely when top-2 and top-3 are taken into account, the exact matches nearly doubled the score from top-1 to top-3.
However, similar to HiTyper-DL, the majority of the correct predictions are within the top-3 types.
Overall, while ML-based tools demonstrate promise, they still trail behind the performance of \toolname when considering the top-1 predictions.

\begin{table}
	\centering
	
	\caption{top-$n$ exact matches comparison with ML tools	\vspace{-.5em}}
	\label{tab:typeevalpy-table-topn_analysis}
	\renewcommand{\arraystretch}{.8}
	\begin{tabular}{lccccc}
		\toprule
		\textbf{Tool} & \textbf{top-$n$} & \textbf{FR} & \textbf{FP} & \textbf{LV} & \textbf{Total} \\
		\midrule
		HeaderGen & 1 & 186 & 56 & 322 & 564 \\ \midrule
		HiTyper-DL & 1 & 163 & 27 & 179 & 369 \\
		& 3 & 173 & 37 & 225 & 435 \\
		& 5 & 175 & 37 & 229 & 441 \\ \midrule
		Type4Py & 1 & 39 & 19 & 99 & 157 \\
		& 3 & 103 & 31 & 167 & 301 \\
		& 5 & 109 & 31 & 174 & 314 \\
		\bottomrule
	\end{tabular}
	\vspace{-.5em}
\end{table}

\section{Discussion}
\label{sec:discussion}
In this section, we highlight the outcomes for each tool, encapsulating their strengths and weaknesses as observed by \typeevalpy:

\textbf{HeaderGen.}
The analysis revealed that \toolname performed consistently across all categories.
With flow-sensitive analysis built on top of PyCG, \toolname is able to infer types of Python elements accurately.
However, it showed limitations in the \texttt{builtins} and \texttt{external} categories, highlighting the lack of support for analyzing external source code.
The support for utilizing typestubs in analysis is limited, such as, support for typestubs with overriding definitions for the same function based on FP types is not implemented.

\textbf{Jedi.}
The open-source community driven tool Jedi has been enhanced to address diverse challenges and to provide broad analysis capabilities.
Specifically, its ability to reason about external source code in the \texttt{builtins} and \texttt{external} categories is vital for analyzing real-world code.
However, Jedi's design choice to omit the output of types for FPs hinders its overall result.
Furthermore, our analysis of the mismatch reports from \typeevalpy highlighted an inconsistency.
When a function is passed as reference in an argument to a function call, both Jedi and Pyright incorrectly infer the variable type as the return type of that function. In reality, the correct type should be \texttt{callable}, since the function is not actually called.
In \typeevalpy, Pyright and Jedi exhibited this behavior in 57 and 18 instances, respectively, highlighting the potential for improvement.

\textbf{Pyright.}
Microsoft's Pyright demonstrated strong performance in the \texttt{builtins} and \texttt{external} categories. 
It also performed slightly better in the FP and LV categories than Jedi. 
It is also pertinent to mention that, Pyright, developed in TypeScript, lacks interfaces to access its internal analysis structures.
This limitation makes general-purpose analysis difficult.
To address this, we created a language server protocol (LSP) client in Python, allowing \typeevalpy to query results for each element.
The static analysis community can benefit from such an interface if its built into Pyright.

\textbf{Scalpel.}
Scalpel exhibited strong performance in aspects related to functions.
However, it does not currently support the output of LV types, which significantly impacted its overall ranking and needs improvement in this area.
For instance, it can annotate up to 155 function return types and 32 function parameters, ranking second among all pure static analysis tools for the two categories. 
Regarding LV, Scalpel only provides six annotations at the second last position. 
Furthermore, it does not handle external library calls, which are common in real-world projects.

\textbf{Type4Py.}
Type4Py uses a deep similarity learning-based technique, meaning that it can only infer types that were seen during the training phase. 
Therefore, Type4Py performs poorly compared to static approaches such as HeaderGen and Pyright. 
Also, Type4Py only learns from identifiers in the method signature and also the usage of FPs and LVs inside the method body. 
This may also explain why Type4Py fails to infer types for some of the categories in the microbenchmark, i.e., classes, dynamic, and exceptions. 
It is also worth mentioning that Type4Py is mostly trained on local variables data and hence it performs relatively better on LV types compared to FR and FP types. 
For better performance, top-5 suggestions from Type4Py should be considered, providing that it performs $k$-nearest neighbor search to find possible type annotations for a given query.

\textbf{HiTyper \& HiTyper-DL.}
HiTyper is a hybrid type inference approach, which combines a deep learning model, i.e., Type4Py with static analysis. 
As expected, it performs better than Type4Py, a pure ML-based approach. In general, 
HiTyper's static inference part seems to be quite imprecise as it is very unsound but more complete than the other baselines. 
On the other hand, HiTyper-DL is more sound but incomplete. 
This can be explained by the fact that it uses type rejection rules to be more precise. However, HiTyper-DL can be expensive to run for large projects considering its hybrid nature.

Though showing promising results, the hybrid paradigm does not show many advances in soundness. 
Only 15 of 154 sound results are brought by the deep learning model, which, however, largely reduces the completeness by 103 code snippets. 

\textbf{Outlook.}
In our study, we found that \toolname performs reliably in several complex scenarios.
Yet, for code that depends significantly on external libraries and has dependable user-defined type stubs, Pyright and Jedi seem more suitable due to their enhanced integration with type stubs.
Additionally, the hybrid HiTyper-DL approach shows potential.
Future research can explore how \toolname might be combined with HiTyper-DL to improve outcomes.

\section{Conclusion}
\label{sec:conclusion}
In this paper, we presented \typeevalpy, a micro-benchmarking framework designed for assessing Python type inference tools.
Our comprehensive analysis covered a diverse array of six type inference tools including static analysis based approaches, ML-based approaches, and hybrid approaches.
Notably, \toolname performed the best in terms of exact matches, soundness, and completeness.
Jedi and Pyright followed close to each other, ranking second and third, respectively.
Moreover, HiTyper-DL, a hybrid type inference tool, demonstrated potential advantages over solely ML-based alternatives, securing the fourth position.
The comparative insights from \typeevalpy highlights the differences between these tools and sets the stage for future research and optimization endeavors.
Overall, the challenge of type inference in Python remains unresolved, presenting opportunities for advancement in terms of both soundness and completeness.



\bibliographystyle{ACM-Reference-Format}
\bibliography{sample-base}

\end{document}